\documentclass[final]{svjour2}
\usepackage{graphicx}
\usepackage{rotating}
\usepackage{amssymb}
\usepackage{mathptmx}
\usepackage[numbers]{natbib}
\makeatletter
\journalname{Journal of Low Temperature Physics}
\bibpunct{}{}{,}{s}{}{,}

\begin{document}

\newcommand{\hdblarrow}{H\makebox[0.9ex][l]{$\downdownarrows$}-}
\title{Time Evolution of Electric Fields in CDMS Detectors}

\author{S.W.~Leman$^1$ \and D. Brandt$^2$ \and P.L.~Brink$^2$ \and B.~Cabrera$^3$ \and H. Chagani$^4$ \and M.~Cherry \and P.~Cushman$^4$ \and E.~Do~Couto~E~Silva$^2$ \and T. Doughty$^5$ \and E.~Figueroa-Feliciano$^1$ \and V. Mandic$^4$ \and K.A.~McCarthy$^1$ \and N.~Mirabolfathi$^5$ \and M.~Pyle$^3$ \and A.~Reisetter$^6$ \and R. Resch$^2$ \and B.~Sadoulet$^5$ \and B.~Serfass$^5$ \and K.M.~Sundqvist$^5$ \and A. Tomada$^2$ \and B.A.~Young \and J. Zhang$^4$ \and on behalf of the Cryogenic Dark Matter Search collaboration}

\institute{1:Massachusetts Institute of Technology, Kavli Institute, \\ Cambridge, MA 02139, USA\\
\email{swleman@mit.edu} \\
2: SLAC National Accelerator Laboratory / KIPAC, \\Menlo Park, CA 94025, USA \\
3: Stanford University, Department of Physics, \\Stanford, CA 94305, USA \\
4: University of Minnesota, Minneapolis, School of Physics \& Astronomy, Minnesota 55455, USA \\
5: The University of California at Berkeley, Department of Physics, \\Berkeley, CA 94720, USA \\
6: St. Olaf College, Department of Physics, Northfield, Minnesota 55057, USA \\
}

\date{07.20.2011}

\maketitle


\begin{abstract}

The Cryogenic Dark Matter Search (CDMS) utilizes large mass, 3" diameter x 1" thick target masses as particle detectors. The target is instrumented with both phonon and ionization sensors, the later providing a $\sim$1~V~cm$^{-1}$ electric field in the detector bulk. Cumulative radiation exposure which creates $\sim$200$\times$10$^6$ electron-hole pairs is sufficient to produce a comparable reverse field in the detector thereby degrading the ionization channel performance. To study this, the existing CDMS detector Monte Carlo has been modified to allow for an event by event evolution of the bulk electric field, in three spatial dimensions. Our most resent results and interpretation are discussed.

\end{abstract}

\section{Introduction}

The Cryogenic Dark Matter Search experiment~\cite{Ahmed2010, Ahmed2011} uses a combination of phonon and ionization signals produced in radiation interactions in the detector target mass to determine event energy and interaction type. Electron recoils, which constitute a background for the experiment, result in 25\% ionization energy with the remaining 75\% energy in the phonon system.  Nuclear recoils, which make up the signal region, result in a reduced ionization signal to $\sim$1/3 the amount for gamma events, with the remaining amount in the phonon system. Cuts are imposed on the ratio of ionization and phonon energy to allow passage of nuclear recoil events.

While not the focus of this paper, events located outside of the bulk fiducial volume result in reduced ionization energy leading to electron recoils appearing like nuclear recoils. One example is surface events in which the high initial kinetic energy of charge carriers causes them to be injected into the wrong electrode. Another example are events at large radius, where fringing fields causes charge carriers to transport into the detector sidewall.

There are also detector operating states which result in reduced ionization energy readout, again causing misidentification of electron recoils as nuclear recoils. This state is described by a build up of space charge in the detector which sets up an electric field opposite to the applied field. This results in a net zero electric field which does not accelerate charges into the readout channel and causes an increased charge trapping rate within the detector bulk.

\section{Charge Transport}

Charge transport in germanium at low temperatures is described in more detail in another paper in these proceedings and elsewhere~\cite{LemanLTD14Orientation, Leman2011_5}. In our CDMS detector Monte Carlo, transport stops when either a charge traps on impurities or reaches the detector-vacuum boundary. While there is variation in trapping times between the CDMS detectors, the Monte Carlo models used in this paper used a representative value deduced from calibration data, 8~$\mu$s and 64~$\mu$s for electron and hole trapping respectively. The carrier stopping location is then recorded and a process of updating the electric potential inside of the detector, to include the initial potential solution and the contribution from these charges begins.

A component of the potential evolution calculation involves solving the electric potential kernel of point charges inside the detector. Boundary conditions that describe the kernel include condition $V=0$ on the lithographically defined metal surfaces and conditions $\epsilon_{out} dV_{out} / dn       -      \epsilon_{in} dV_{in} / dn = - \sigma$ and $dV_{in} / dp = dV_{out} / dp$, where $dn$ and $dp$ represent differential lengths in the normal and parallel directions, at the germanium-vacuum boundary. Due to the complicated shape of the metal surfaces this is a non-trivial kernel to calculate and simpler approximations are used for two different detector types. For the iZIP style detector~\cite{Brink2006}, the top and bottom of the detector have a relatively sparse 6.1\% aluminum area coverage and are spaced $\sim$2~mm away from the neighboring detector resulting in a topology that can be described as a long cylinder of germanium when vacuums are ignored (considered to have a Ge dielectric of $\epsilon=16$). There is a ground boundary condition provided by a copper supporting structure which surrounds the detector. This structure is hexagonal and has a minimum center to copper distance of 38.7~mm, where the detector radius is 38.1~mm. With these and an additional approximation that the copper can is described by a 38.7~mm radius cylinder, a \emph{soup can} kernel is found to be~\cite{Smythe1950}

\begin{equation}
V (\rho, \phi, z) = \sum_{n=0}^\infty  \sum_{r=0}^\infty A_{nr} J_n(k_nr \rho) \cos(n(\phi - \phi_0)) \sinh(k_{nr}(L+z)), ~z \le z_0
\end{equation}
where
\begin{equation}
A_{nr} = \frac{4(2-\delta_{n0})}{a^2}  \frac{\sinh(k_{nr}(L-z_0))}{\sinh(2 k_{nr} L)} \frac{J_n(k_{nr} \rho_0)}{k_{nr}(J_{n+1}(k_{nr}a))^2},
\end{equation}
and the Bessel function $J_n$ is defined to have zeros at the cylinder walls ($r=a$). The kernel below the source is found by interchanging $+$ and $-$ in the $\sinh$ function argument for both $V$ and $A$. 

A high number of $n$ terms constrains the solution in the angular direction and a high number of $r$ terms constrains the solution in the radial and longitudinal directions. Taking the sums to infinity (or some large number) is computationally expensive given the ~200k mesh points and ~30 charge carrier pairs. Instead, the sums are computed to $n=3$ and $r=81$ for the iZIP detector resulting in a spread of $\sim \pi /2$ in the angular direction and $\sim$2~mm in the radial and longitudinal directions. This spread is reasonable since the iZIP data was exposed to a high energy barium-131 gamma source (356~keV) resulting in events distributed throughout the detector bulk. With the exception of $A_{nr}$, all components in $V$ are independent of charge location and for computational efficiency are computed in an initialization procedure and stored in memory for reuse. Charge carriers in the next event are then propagated with this updated potential solution and the procedure is repeated.

\section{iZIP Monte Carlo Results and Comparison to Data}

In addition to updating the electric potential, the charge carrier stopping position can also be run through an instrumentation simulator to recreate the detector response and allow a direct comparison between Monte Carlo and data. There are some measurement quantities which show qualitative similarities between Monte Carlo and data and some which do not show good agreement.

\begin{figure}[t]
\begin{tabular}{c}
\begin{minipage}{0.48\hsize}
\begin{center}
\includegraphics[width=6cm]{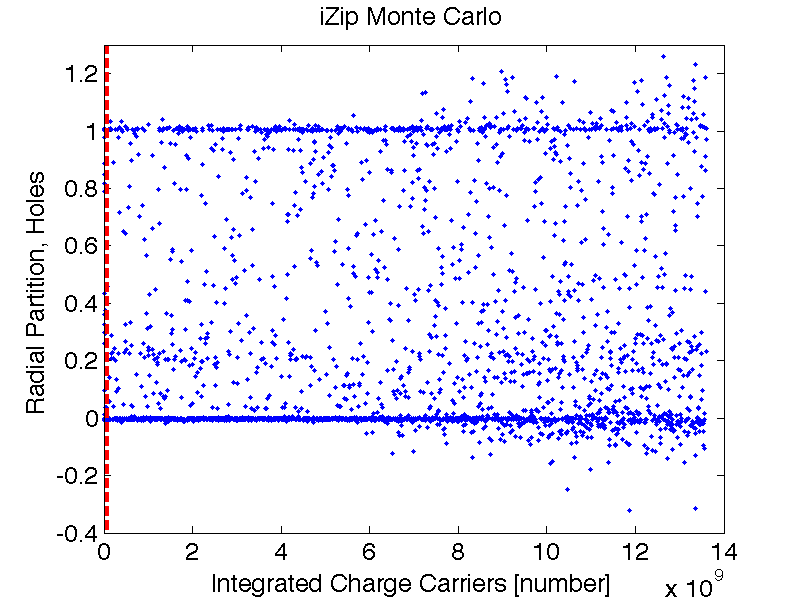}
\end{center}
\end{minipage}
\begin{minipage}{0.48\hsize}
\begin{center}
\includegraphics[width=6cm]{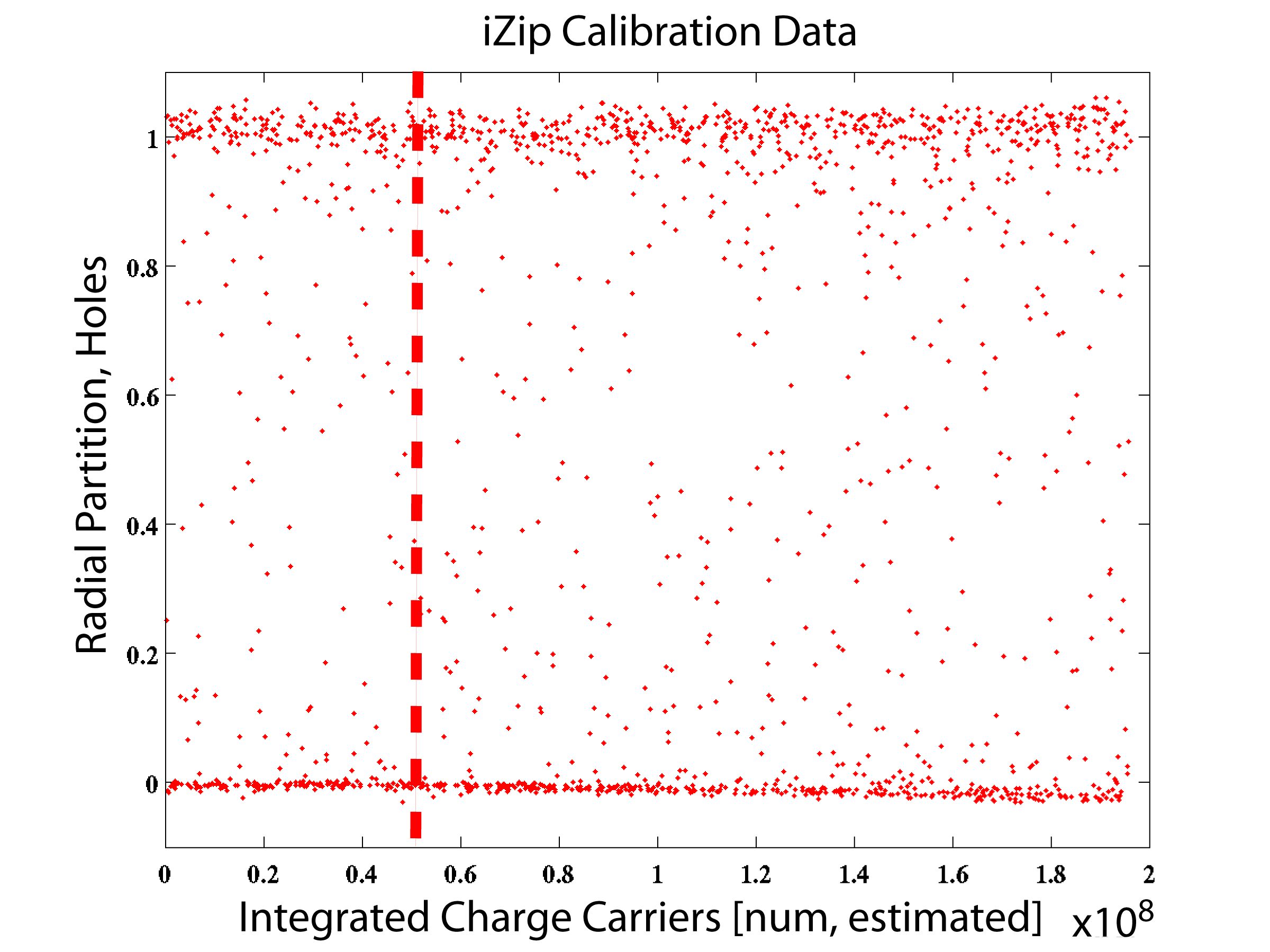}
\end{center}
\end{minipage}
\end{tabular}
\caption[] { \label{fig:QrHole} Radial charge partition from holes as measured in a CDMS iZIP detector. The figure on the left shows a Monte Carlo response of the detector whereas the one on the right is from calibration data. The vertical dashed line indicates periods of similar estimated exposure.}
\end{figure}

An indicator of good agreement is seen in the hole radial-charge-partition quantities for the iZIP detector. This signal is defined to be the signal in the inner electrode minus the signal in the high radius electrode. As shown in Figure~\ref{fig:QrHole} both data and Monte Carlo show this measurement becoming negative as the field evolution progresses indicating a build up of holes at large radius near the hole collection electrode. 

\begin{figure}[t]
\begin{tabular}{c}
\begin{minipage}{0.48\hsize}
\begin{center}
\includegraphics[width=6cm]{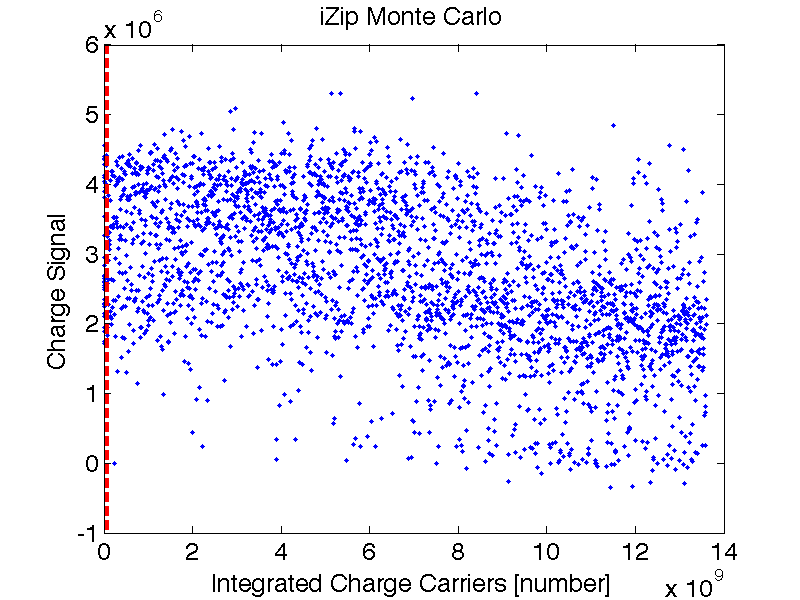}
\end{center}
\end{minipage}
\begin{minipage}{0.48\hsize}
\begin{center}
\includegraphics[width=6cm]{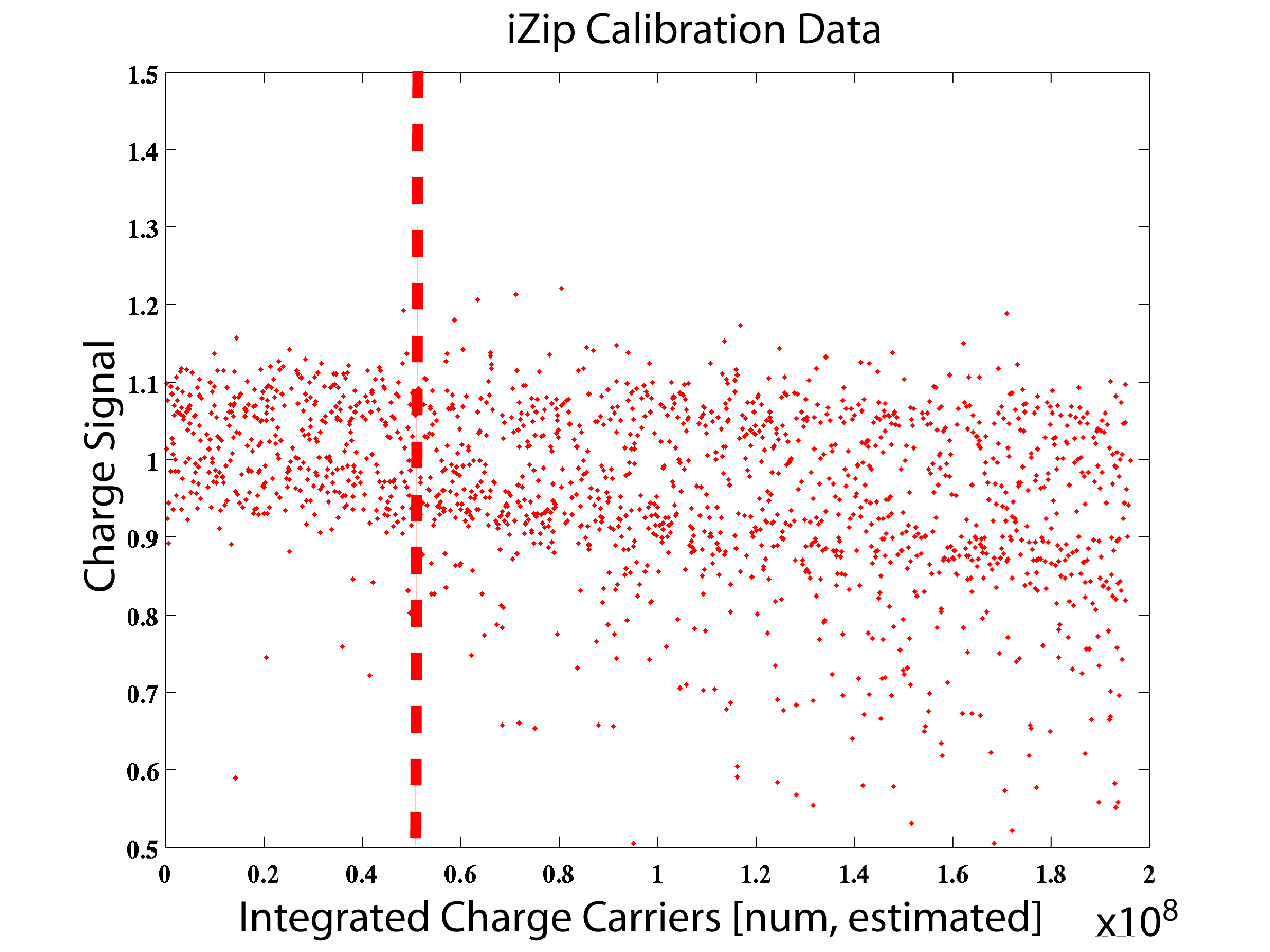}
\end{center}
\end{minipage}
\end{tabular}
\caption[] { \label{fig:Qsum} Charge signal as measured in a CDMS iZIP detector. The figure on the left shows a Monte Carlo response of the detector whereas the one on the right is from calibration data. The vertical dashed line indicates periods of similar estimated exposure.}
\end{figure}

A second indicator of good agreement is seen in the sum of electron and hole collection in the iZIP detector. As shown in Figure~\ref{fig:Qsum} both Monte Carlo and data have a reduced response at long times. A  similar feature is seen in more detailed studies described in a companion paper in these proceedings.~\cite{DoughtyLTD14} This is generally assumed to be due to an induced electric field from trapped charges counteracting the applied electric field resulting in a reduced electric field magnitude in the detector. With this reduced electric field, the charges transport more slowly through the detector increasing their chance of becoming trapped and resulting in a reduced charge collection signal.

\begin{figure}[t]
\begin{tabular}{c}
\begin{minipage}{0.48\hsize}
\begin{center}
\includegraphics[width=6cm]{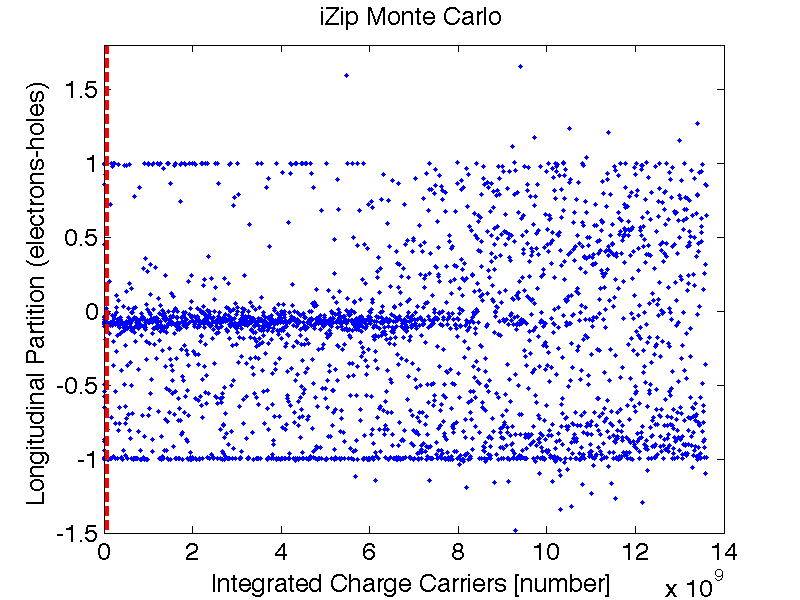}
\end{center}
\end{minipage}
\begin{minipage}{0.48\hsize}
\begin{center}
\includegraphics[width=6cm]{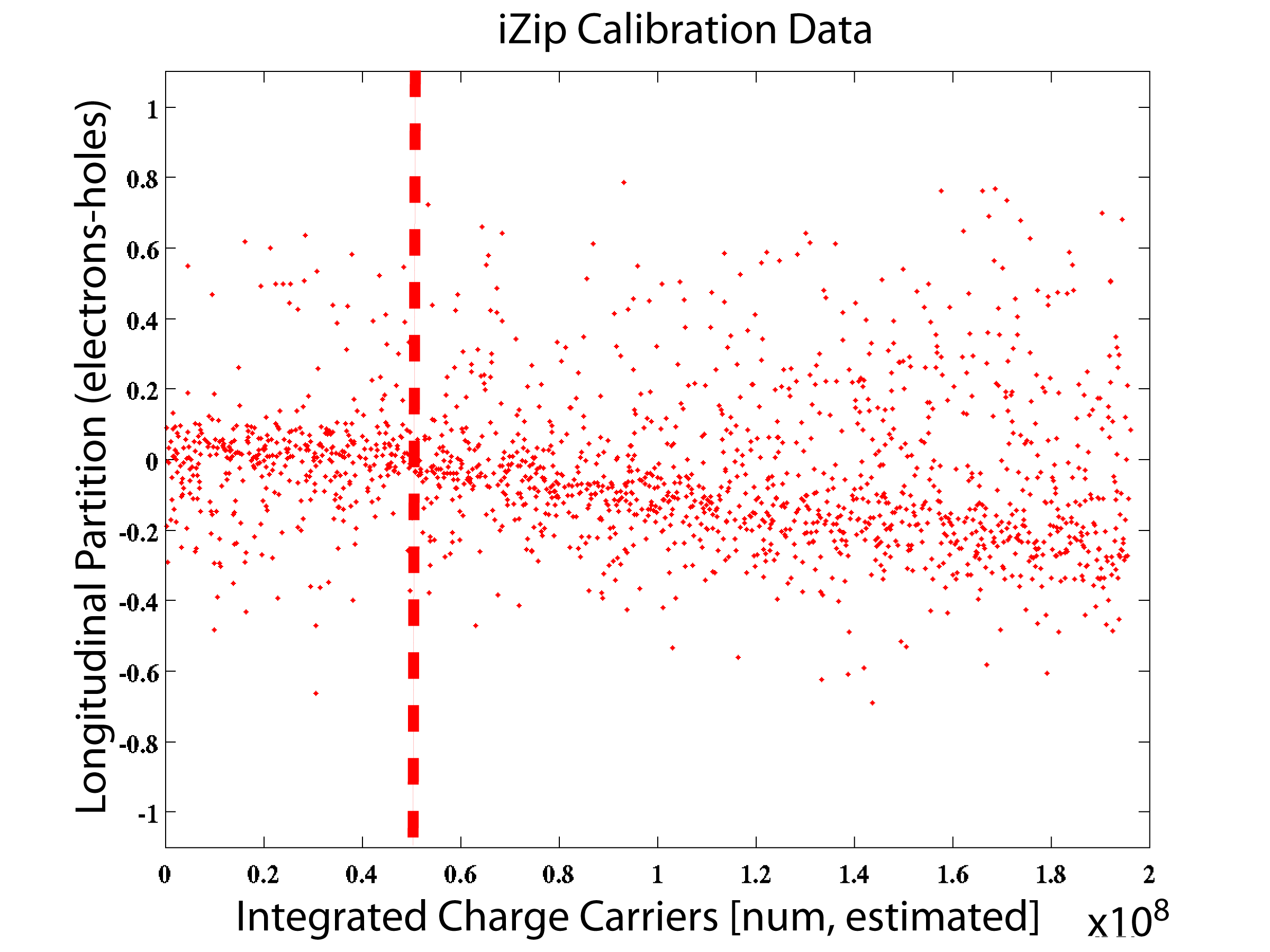}
\end{center}
\end{minipage}
\end{tabular}
\caption[] { \label{fig:Qz} Longitudinal charge partition as measured in a CDMS iZIP detector. The partition is defined to be the hole signal minus the electron signal. The figure on the left shows a Monte Carlo response of the detector whereas the one on the right is from calibration data. The vertical dashed line indicates periods of similar estimated exposure.}
\end{figure}

An indicator of poor agreement between Monte Carlo and data is seen in the charge partitioning along the longitudinal direction of the iZIP detector.  The data in Figure~\ref{fig:Qz} shows a gradual decline from 0 to -0.1 indicating that electrons are transporting less efficiently to the charge electrode as the run progresses. This gradual decrease is strikingly absent in the Monte Carlo and instead there is a dramatic reduction of bulk events (partition = 0) and the measurement becomes increasingly noisy. 

A second indicator of poor agreement is observed in all of the measurements, notably the time scale for any change in measurement are about a factor of 100 greater than what is seen in data. This suggests one of a few possibilities, that the detector was not initially in applied-field only condition when the dataset acquisition was started, that some other mechanism other than gamma-ray interactions is generating charge carriers which reduce the electric field, or that the charge trapping rates are incorrect. In Figures~\ref{fig:QrHole}, \ref{fig:Qsum} and~\ref{fig:Qz}, the number of charge carriers in the Monte Carlo is known, however the number of carriers in data is inferred from the energy of events with energy $>$40~keV. It is additionally possible that events could have been created by lower energy, unresolvable events.

\section{Test Device Monte Carlo Results and Comparison to Data}

In addition to the iZIP detector, a charge transport test device with four concentric ionization channels patterned on one detector face was run. This test device is 100~mm in diameter and 33.3~mm thick and the top and bottom surfaces have a high $\sim$100\% area coverage. Additional discussion of this device and operating characteristics is found in these proceedings.~\cite{ChaganiLTD14}This detector is also surrounded by a hexagonal copper can with minimum center to can distance of 51.5~mm. This detector was exposed to Am-241 gammas (59.5~keV) from four collimated, surface sources. Since the sources (and therefore events) are more localized it is preferable to more tightly constrain the kernel in the angular direction and sums are computed out to $n=27$ and $r=81$ constraining the solution to $\sim$5~mm in the angular direction. 

\begin{figure}[t]
\begin{tabular}{c}
\begin{minipage}{0.48\hsize}
\begin{center}
\includegraphics[width=6cm]{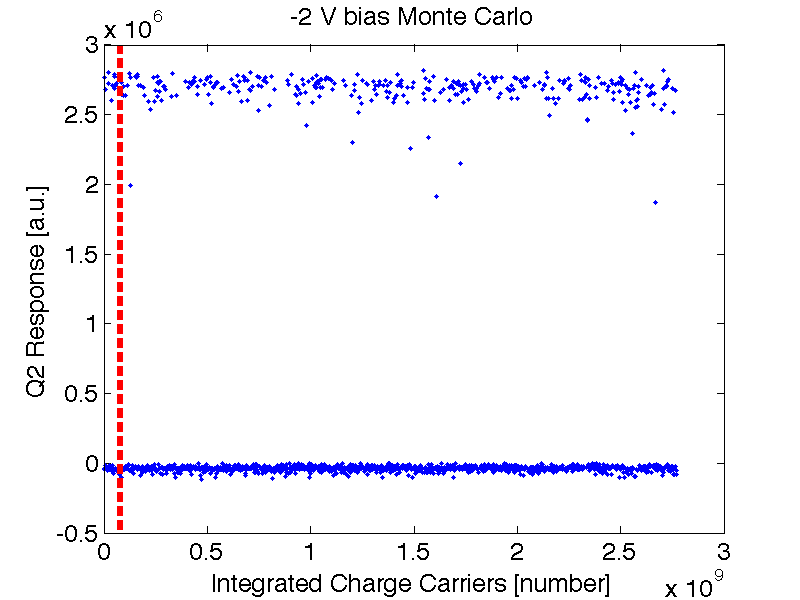}
\end{center}
\end{minipage}
\begin{minipage}{0.48\hsize}
\begin{center}
\includegraphics[width=6cm]{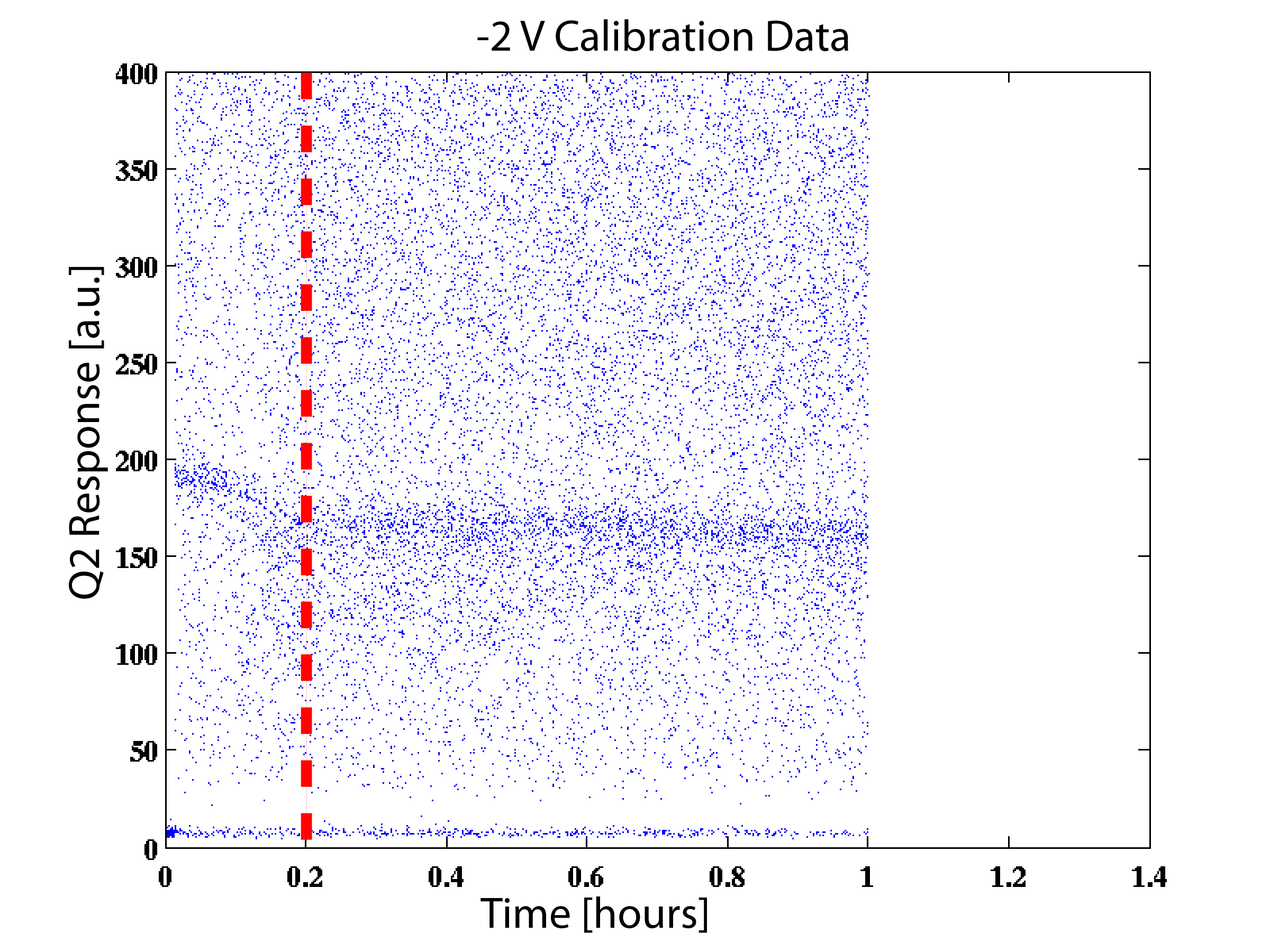}
\end{center}
\end{minipage}
\end{tabular}
\caption[] { \label{fig:G102MC} Response of ionization test device to Am-241 gamma rays. The figure on the left shows a Monte Carlo response of the detector whereas the one on the right is from calibration data. The lines at high value, initially at $\sim$200 in data are from gammas originating in the collimated source above the Q2 electrode. The vertical dashed line indicates periods of similar estimated exposure. The Monte Carlo results show an additional line with response $\sim$0 for events generated above one of the other three electrodes.}
\end{figure}

The comparison between Monte Carlo and data do not show good agreement as seen in Figure~\ref{fig:G102MC}. Shown in these figures is the response of the second to innermost electrode to Am-241 gamma rays. The Monte Carlo never shows a change in charge response despite being run for an order of magnitude longer than the charge degradation time scale seen in calibration data. 

Charges that stop at or near a detector face would result in a large amount of image charges in the metal film on the detector face. This results in a kernel in which charges stopped near a detector face have a small effect on electric field orientation. This would suggest that most of the degradation in the test device is due to charges trapping in the detector bulk and the lack of degradation in Monte Carlo is due to a charge trapping time that is too small for this detector.

\section{Conclusions and Future Work}

There is possible qualitative agreement between MC and calibration data in the iZIP radial charge partition and overall signal but poor agreement in longitudinal charge partition and the time scale for degradation in the iZIP and ionization test device. The Monte Carlo described in this paper did not take into account the increase in charge trapping rates at low fields and including this may improve the Monte Carlo results. Furthermore, the charge trapping times may be significantly higher in the ionization test device compared to what was included in Monte Carlo and this additional improvement in Monte Carlo may improve comparison.

\begin{acknowledgements}

This work is supported in part by the United States National Science Foundation (under Grant Nos. PHY-0847342, 0705052 , 0902182 and 1004714) and the United States Department of Energy (under contract DE-AC02-76SF00515).

\end{acknowledgements}

\pagebreak


\begin{thebibliography}{99}

\bibitem{Ahmed2010} Z. Ahmed et al. Science 327:1619-1621 (2010)

\bibitem{Ahmed2011} Z. Ahmed et al, PRL, {\bf 106}, 131302 (2011)

\bibitem{LemanLTD14Orientation} S.W. Leman et al, these proceedings

\bibitem{Leman2011_5} S.W. Leman, Review of Physics and Monte Carlo Techniques as Relevant to a Cryogenic, Phonon and Ionization Readout, Radiation-Detector, submitted 2011, Review of Modern Physics

\bibitem{Brink2006} P.L. Brink et al, Nuclear Instruments and Methods in Physics Research, A559 (2006) 414Ð416

\bibitem{Smythe1950}  W.R. Smythe, Static and Dynamic Electricity, McGraw-Hill Book Company, Inc., NY, 1950

\bibitem{DoughtyLTD14} T. Doughty et al, these proceedings

\bibitem{ChaganiLTD14} H. Chagani et al, these proceedings


\end{thebibliography}
\end{document}